\documentstyle{article}
\begin{document}
\setlength{\baselineskip}{15pt}
\title{The range of elementary interactions: a criticism of the standard 
classroom approach.}
\author{ Benito Hern\'{a}ndez--Bermejo }
\date{}
\maketitle

{\em Departamento de F\'{\i}sica Fundamental, Facultad de Ciencias, 
Universidad Nacional de Educaci\'{o}n a Distancia. Apartado 60.141, 28080 
Madrid (Spain). } 

\mbox{}

\mbox{}

\begin{center}
{\bf Abstract}
\end{center}
The finite scope of the some elementary interactions is usually presented to 
Physics students as a natural consequence of the time-energy uncertainty 
relation. It is demonstrated that this heuristic derivation is not {\em a 
priori\/} valid. Accordingly, it is argued the doubtful usefulness of this 
``proof'' in the classroom, for it may lead to misconceptions. 

\mbox{}

\mbox{}

\begin{center}
{\bf Resumen}
\end{center}
Usualmente se presenta a los estudiantes de F\'{\i}sica el alcance finito de 
ciertas interacciones fundamentales como una consecuencia natural de la 
relaci\'{o}n de incertidumbre energ\'{\i}a-tiempo. Se demuestra que esta 
derivaci\'{o}n heur\'{\i}stica no es {\em a priori\/} v\'{a}lida. Por tanto, 
se sostiene que es dudosa la utilidad de una tal ``demostraci\'{o}n'' en el 
aula, dado que puede conducir a concepciones err\'{o}neas. 

\mbox{}

\mbox{}

\mbox{}


\mbox{}

\mbox{}

Short title: Range of interactions.

\pagebreak

In many textbooks, both introductory (see for example \cite{alf}) and 
advanced (such as \cite{kra,ham}, to cite a few) the finite range of an 
elementary interaction with massive field quanta is explained as follows: 
assume that the intermediary particle of a force possesses mass $m$. Then, 
the energy required to produce it is $\Delta E = mc^2$. Thus, from 
Heisenberg's principle the life of the particle will be of order 
\begin{equation}
   \Delta t = \frac{ \hbar }{ 2 \; \Delta E } = \frac{ \hbar }{ 2mc^2 }
\end{equation}
Since an upper bound for the speed of the virtual particle is $c$, an upper 
bound for the range $l$ of the interaction will be 
\begin{equation}
   l_{max} = c \; \Delta t = \frac{ \hbar }{ 2mc }
   \label{mat}
\end{equation}
Sometimes \cite{ham} this line of reasoning is extended to include also the 
electromagnetic interaction: in this case $\Delta E$ is the energy of the 
exchanged virtual photon, and from a straightforward repetition of the 
previous argument we obtain: 
\begin{equation}
   l_{max} = c \; \Delta t = \frac{ c \hbar }{ 2 \; \Delta E }
   \label{luz}
\end{equation}
In either case, $l_{max}$ should be an estimation of the {\em maximum\/} 
distance that the virtual particle can travel before being reabsorbed. 

However, this is a misleading argument. To see this, it should be noticed 
that the time-energy uncertainty relation is \cite{coh,gap}: 
\begin{equation}
   \Delta E \; \Delta t \geq \frac{ \hbar }{ 2 }
   \label{up}
\end{equation}
Therefore, from this equation it is only possible to deduce a {\em lower\/} 
bound for $\Delta t$. Then, if $v < c$ is the average speed of the virtual 
particle, we arrive at: 
\begin{equation}
   l = v \; \Delta t \geq \frac{ v \hbar }{ 2 \; \Delta E }
   \label{lb}
\end{equation}
Hence, the time-energy uncertainty relation implies the existence of a 
{\em core\/} of radius $v \hbar / ( 2 \; \Delta E)$ such that the virtual 
particle cannot be reabsorbed inside it. In other words, equation (\ref{lb}) 
establishes a {\em lower\/} bound for the range of the interaction. 

To retrieve a formula similar to (\ref{mat}) or (\ref{luz}), it is first 
necessary to assume that the actual value of $l$ is of the order of the lower 
bound in (\ref{lb}): 
\begin{equation}
   l \simeq l_{min} = \frac{ v \hbar }{ 2 \; \Delta E }
   \label{mis}
\end{equation}
Notice that this hypothesis is extraneous to the uncertainty principle 
itself. Then, since $v < c$, we obviously have:
\begin{equation}
   l \simeq l_{min} \leq \frac{ c \hbar }{ 2 \; \Delta E }
   \label{kkf}
\end{equation}
This statement is apparently similar to that expressed by equation 
(\ref{luz}) (or by (\ref{mat}), if we substitute $\Delta E$ by $mc^2$). 
However, the meaning of (\ref{kkf}) is entirely different to this one, since 
in (\ref{kkf}) we have an upper bound {\em for a lower bound,\/} that is, 
an upper bound for the radius of the core. Clearly, the obtainment of the 
standard heuristic formula requires an awkward argument, and the 
interpretation of the result looks too involved and confusing for the 
student.

Consequently, the previous development seems to imply that the usual 
heu\-ristic derivation is undesirable from a pedagogic point of view. As we 
have seen, it does not provide a complete description of the process. 
Moreover, to explain the finite range of the interaction we are in fact 
making use of an argument leading to the opposite result: the existence of a 
lower bound for the range or, in other words, the existence of a core where 
the interaction is forbidden. Conversely, as the author has had the 
opportunity to observe, the well-known fact that the life 
of virtual particles is finite may lead to the student to {\em deduce\/} that 
the uncertainty principle is $\Delta t \leq \hbar / ( 2 \; \Delta E )$ 
instead of the correct expression (\ref{up}), because the wrong formula leads 
inmediately to the expected result in which $l$ has an upper bound given by: 
\begin{equation}
l_{max} = c \; \Delta t \leq \frac{ c \hbar }{ 2 \; \Delta E }
\end{equation}

It seems therefore reasonable to conclude that the standard trick used to 
establish the finite range of those interactions with massive field quanta 
should be carefully employed, or perhaps avoided, in the classroom. 

\pagebreak

\end{document}